\documentclass[epj]{webofc}
\usepackage[varg]{txfonts}   
\wocname{EPJ Web of Conferences}
\woctitle{CONF12}
\begin{document}
\selectlanguage{english}
\title{Flavour decomposition of electromagnetic transition form \\ factors of 
nucleon resonances}
%
%

\author{
Jorge Segovia\inst{1}\fnsep\thanks{\email{jorge.segovia@tum.de}} \and
Craig D. Roberts\inst{2}
}

\institute{
Physik-Department, Technische Universit\"at M\"unchen,
James-Franck-Str.\,1, D-85748 Garching, Germany 
\and
Physics Division, Argonne National Laboratory, Argonne, Illinois
60439, USA
}

\abstract{%
In Poincar\'e-covariant continuum treatments of the three valence-quark 
bound-state problem, the force behind dynamical chiral symmetry breaking also 
generates nonpointlike, interacting diquark correlations in the nucleon and its 
resonances. We detail the impact of these correlations on the nucleon's elastic 
and nucleon-to-Roper transition electromagnetic form factors, providing 
flavour-separation versions that can be tested at modern facilities.
}
\maketitle


\section{Introduction}
\label{intro}

Dynamical chiral symmetry breaking (DCSB) is a theoretically-established 
feature of Quantum Chromodynamics (QCD) and the most important mass generating 
mechanism for visible matter in the Universe, being responsible for 
approximately $98\%$ of the proton's mass. A fundamental expression of DCSB is 
the behaviour of the quark mass-function, $M(p)$. This appears in the 
dressed-quark propagator which may be obtained as a solution to the most famous 
and simple QCD's Dyson-Schwinger equation: the gap 
equation~\cite{Cloet:2013jya}. The nontrivial character of 
the mass function arises primarily because a dense cloud of gluons comes to 
clothe a low-momentum quark. It explains how an almost-massless parton-like 
quark at high energies transforms, at low energies, into a constituent-like 
quark with an effective mass of around $350\,{\rm MeV}$.

DCSB ensures the existence of nearly-massless pseudo-Goldstone modes (pions), 
but another equally important consequence of DCSB is less well known. Namely, 
any interaction capable of creating pseudo-Goldstone modes as bound-states of a 
light dressed-quark and -antiquark, and reproducing the measured value of their 
leptonic decay constants, will necessarily also generate strong 
colour-antitriplet correlations between any two dressed quarks contained within 
a baryon. Although a ri\-go\-rous proof within QCD cannot be claimed, this 
assertion is based upon an accumulated body of evidence, gathered in two 
decades of studying two- and three-body bound-state problems in hadron 
physics (the interested reader is referred to the discussion in 
Ref.~\cite{Segovia:2015ufa} and to Refs.~[21-35] cited therein). No realistic 
counter examples are known; and the existence of such diquark correlations is 
also supported by simulations of lattice-regularised 
QCD~\cite{Alexandrou:2006cq, Babich:2007ah}.

The diquark correlations predicted to exist within baryons are not the static, 
pointlike degrees-of-freedom which were historically introduced in order to 
solve the fact that many states predicted within the $SU(6)\otimes O(3)$ quark 
model were missing from the observed spectrum. The modern dynamical diquark 
correlations are nonpointlike and fully interacting, and each of a baryon's 
three dressed-quarks is involved in every type of correlation to the fullest 
extent allowed by its quantum numbers and those of the bound state.  One should 
therefore expect the spectrum obtained in the presence of such correlations to 
be as rich as that allowed by a three-constituent quark 
model~\cite{Eichmann:2016yit, Eichmann:2016hgl}.

It has been argued that the presence of diquark correlations should have 
observable consequences on the elastic and transition form factors of nucleon 
resonances~\cite{Segovia:2015ufa, Segovia:2016zyc}. Many related insights have 
been revealed in a series of recent articles~\cite{Segovia:2013rca, 
Segovia:2013uga, Segovia:2014aza, Segovia:2015hra} focused on the calculation of 
the Nucleon, Delta and Roper elastic and transition form factors using a 
widely-used leading-order (rainbow-ladder) truncation of QCD's Dyson-Schwinger 
equations and comparing results between a QCD-based framework and a 
vector$\,\otimes\,$vector contact interaction. It is our purpose here reviewing 
some of the outcomes of such studies, paying particular attention to those 
directly related with strong diquark correlations and can be tested at modern 
facilities.


\begin{figure}[!t]
\begin{center}
\hspace*{0.50cm}
\includegraphics[clip,width=0.40\textwidth,height=0.18\textheight]
{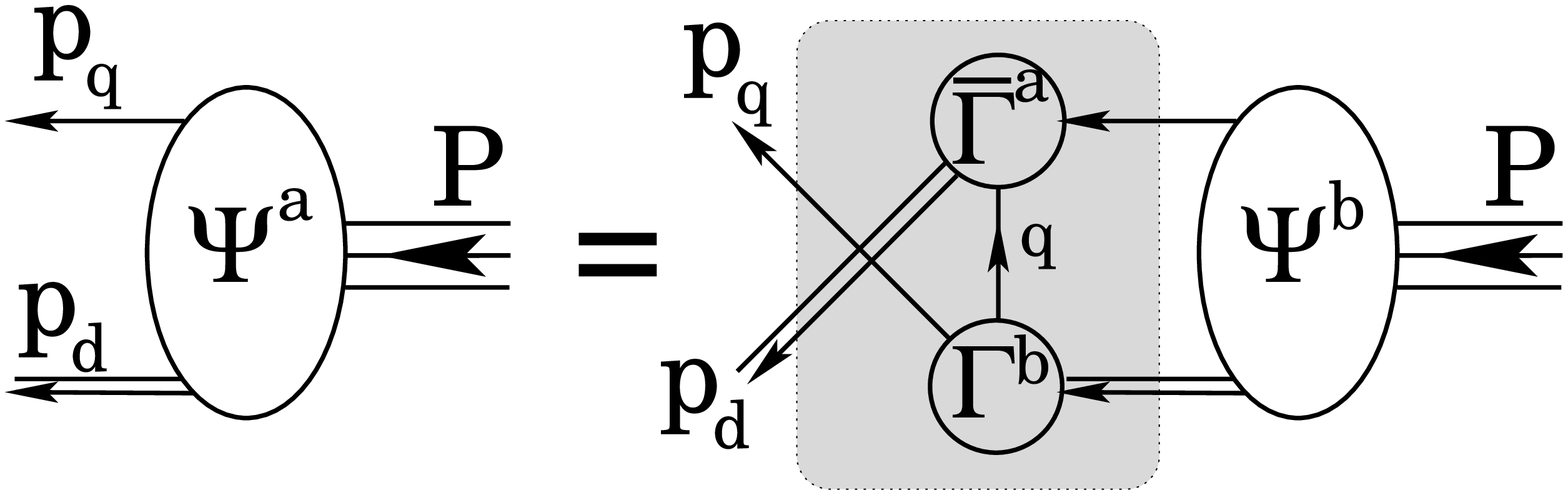} 
\hspace*{1.00cm}
\includegraphics[clip,width=0.40\textwidth,height=0.20\textheight]
{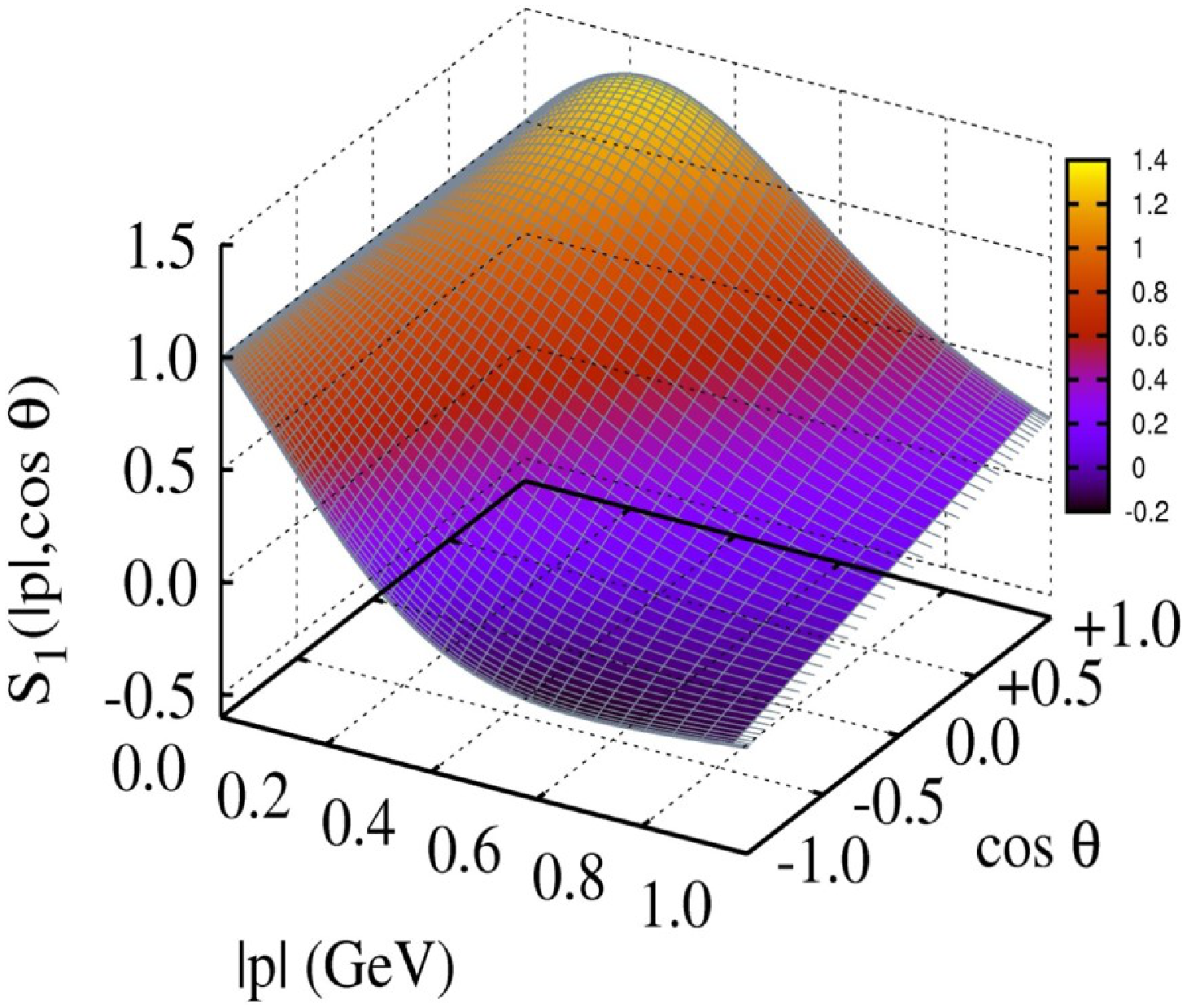}
\caption{\label{fig:Faddeev} {\it Left panel:} Poincar\'e covariant 
Faddeev equation. $\Psi$ is the Faddeev amplitude for a baryon of total 
momentum $P= p_q + p_d$, where $p_{q,d}$ are, respectively, the momenta of the 
quark and diquark within the bound-state. The shaded area demarcates the 
Faddeev equation kernel: {\it single line}, dressed-quark propagator; $\Gamma$, 
diquark correlation amplitude; and {\it double line}, diquark propagator.
{\it Right panel:} Dominant piece in the nucleon's eight-component 
Poincar\'e-covariant Faddeev amplitude: $S_1(|p|,\cos\theta)$. In the nucleon 
rest frame, this term describes that piece of the quark--scalar-diquark 
relative momentum correlation which possesses zero intrinsic quark--diquark 
orbital angular momentum, i.e. $L=0$, before the propagator lines are 
reattached to form the Faddeev wave function. Referring to the right panel of
Fig.~\ref{fig:Faddeev}, $p= P/3-p_q$ and $\cos\theta = p\cdot P/\sqrt{p^2 
P^2}$. The amplitude is normalised such that its $U_0$ Chebyshev moment is 
unity at $|p|=0$, see Ref.~\cite{Segovia:2014aza}.
}
\end{center}
\end{figure}

\section{Baryon structure and the nucleon's form factors}
\label{sec:Baryons}

In the quark$+$diquark picture, baryons are described by the Poincar\'e 
covariant Faddeev equation depicted in the left panel of Fig.~\ref{fig:Faddeev}. 
Two main contributions appear in the binding energy: i) the formation of tight 
diquark correlations and ii) the quark exchange depicted in the shaded area of 
the left panel of Fig.~\ref{fig:Faddeev}\footnote{Whilst an explicit three-body 
term might affect fine details of baryon structure, the dominant effect of 
non-Abelian multi-gluon vertices is expressed in the formation of diquark 
correlations~\cite{Eichmann:2009qa}.}. This exchange ensures that diquark 
correlations within the baryon are fully dynamical: no quark holds a special 
place because each one participates in all diquarks.

The quark$+$diquark structure of the nucleon is elucidated in the right panel 
of Fig.~\ref{fig:Faddeev}, which depicts the leading component of its Faddeev 
amplitude: $S_1(|p|,\cos\theta)$~\cite{Segovia:2014aza}. This function describes 
a piece of the quark$+$scalar-diquark relative momentum correlation. Notably, in 
this solution of a realistic Faddeev equation there is strong variation with 
respect to both arguments. Support is concentrated in the forward direction, 
$\cos\theta >0$, so that alignment of $p$ and $P$ is favoured; and the 
amplitude peaks at $(|p|\simeq m_N/6,\cos\theta=1)$, whereat $p_q \sim p_d \sim 
P/2$ and hence the natural relative momentum is zero. In the antiparallel 
direction, $\cos\theta<0$, support is concentrated at $|p|=0$, {\it i.e.} $p_q 
\sim P/3$, $p_d \sim 2P/3$. A realistic nucleon amplitude is evidently a 
complicated function; and significant structure is lost if simple interactions 
and/or truncations are employed in building the Faddeev kernel, \emph{e.g}.\ 
extant treatments of a momentum-independent quark-quark interaction -- a contact 
interaction -- produce a Faddeev amplitude that is also momentum independent 
\cite{Wilson:2011aa, Cloet:2014rja}, a result exposed as unrealistic by the 
right panel of Fig.~\ref{fig:Faddeev} for any probe sensitive to the nucleon 
interior.

\begin{figure}[!t]
\centerline{
\begin{tabular}{ll}
\hspace*{-0.50cm}
\includegraphics[clip,width=0.425\textwidth,height=0.275\textheight]
{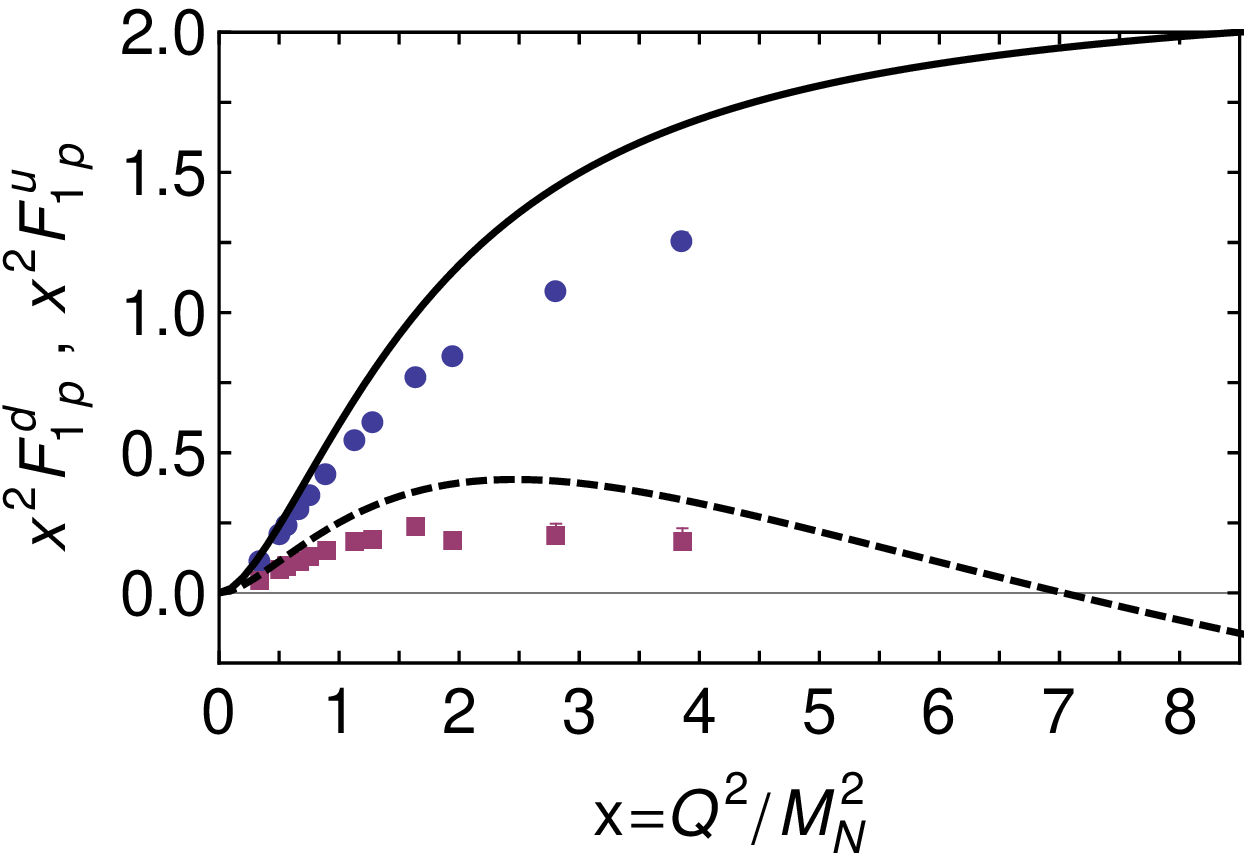} 
&
\hspace*{+0.50cm}
\includegraphics[clip,width=0.425\textwidth,height=0.275\textheight]
{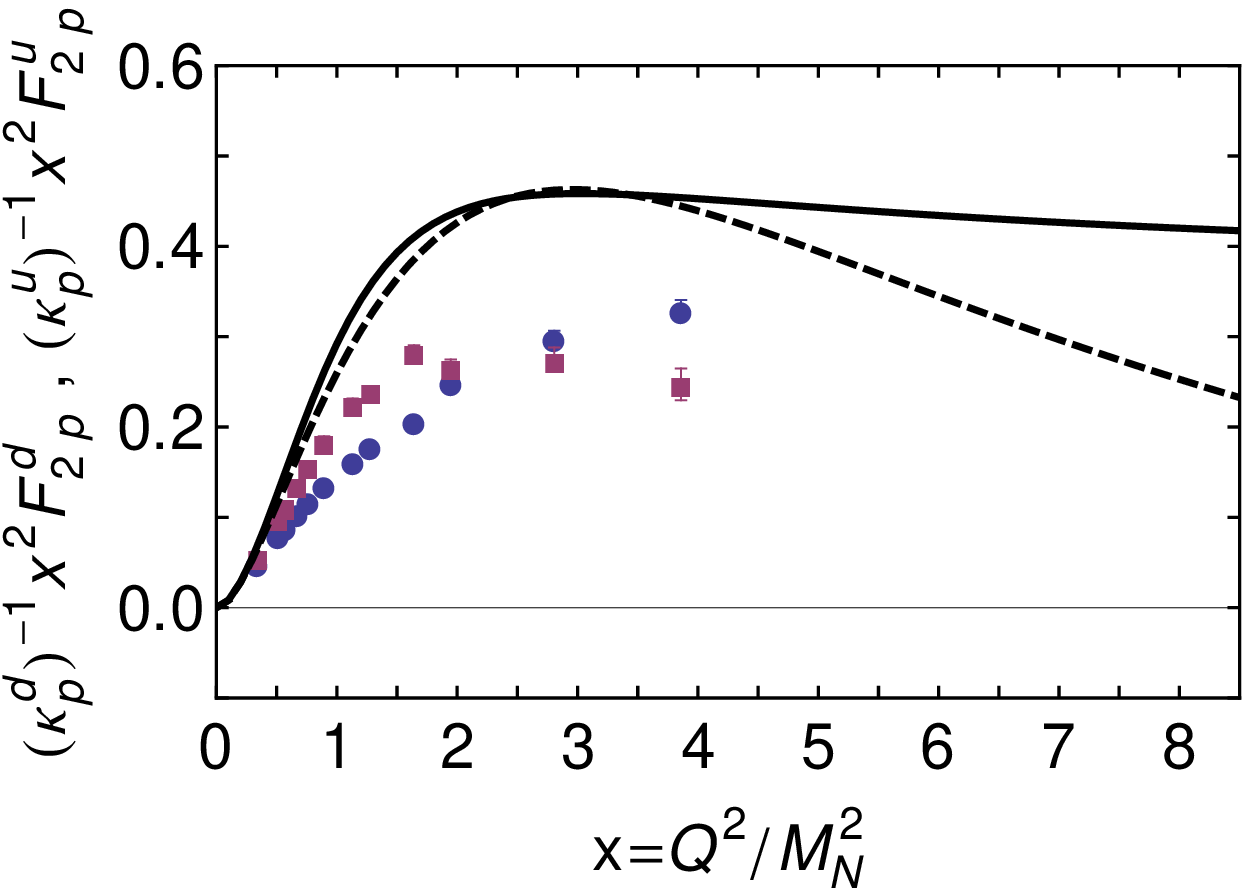}
\end{tabular}
}
\caption{\label{fig:NucNuc_Flavour} {\it Left panel:} Flavour separation of the 
proton's Dirac form factor as a function of $x=Q^2/M_N^2$. The results have 
been obtained using a framework built upon a Faddeev equation kernel and 
interaction vertices that possess QCD-like momentum dependence. The solid-curve 
is the $u$-quark contribution, and the dashed-curve is the $d$-quark 
contribution. Experimental data taken from Ref.~\protect\cite{Cates:2011pz} and 
references therein: circles -- $u$-quark; and squares -- $d$-quark. {\it Right 
panel:} Same for Pauli form factor. We define $\kappa_{u}=F_{2p}^{u}(0)$ and 
$\kappa_{d}=F_{2p}^{d}(0)$ and they can be expressed in terms of the proton 
$(\kappa_{p})$ and neutron $(\kappa_{n})$ anomalous magnetic 
moments~\cite{Cates:2011pz,Segovia:2015ufa}.
%
}
\end{figure}

The presence of diquark correlations inside the nucleon (and its resonances) 
must be evident in numerous empirical differences between the response of the 
bound-state's doubly- and singly-represented quarks to any probe whose 
wavelength is small enough to expose the diquarks' nonpointlike character. In 
connection with electromagnetic probes, Fig.~\ref{fig:NucNuc_Flavour} displays 
the proton's flavour separated Dirac and Pauli form factors. The salient 
features of the data are: the $d$-quark contribution to $F_1^p$ is far smaller 
than the $u$-quark contribution; $F_2^d/\kappa_d>F_2^u/\kappa_u$ on $x<2$ but 
this ordering is reversed on $x>2$; and in both cases the $d$-quark 
contribution falls dramatically on $x>3$ whereas the $u$-quark contribution 
remains roughly constant.

It is natural to seek an explanation for the pattern of behaviour in 
Fig.~\ref{fig:NucNuc_Flavour}. The proton contains scalar and pseudovector 
diquark correlations. The dominant piece of its Faddeev wave 
function is $u[ud]$; namely, a $u$-quark in tandem with a $[ud]$ scalar 
correlation, which produces $62\%$ of the proton's normalisation. If this were 
the sole component, then photon--$d$-quark interactions within the proton would 
receive a $1/x$ suppression on $x>1$, because the $d$-quark is sequestered in a 
soft correlation, whereas a spectator $u$-quark is always available to 
participate in a hard interaction. At large $x=Q^2/m_N^2$, therefore, scalar 
diquark dominance leads one to expect $F^d \sim F^u/x$.  Available data are 
consistent with this prediction but measurements at $x>4$ are necessary for 
confirmation.

\begin{figure}[!t]
\centerline{%
\includegraphics[clip,width=0.45\textwidth,height=0.275\textheight]
{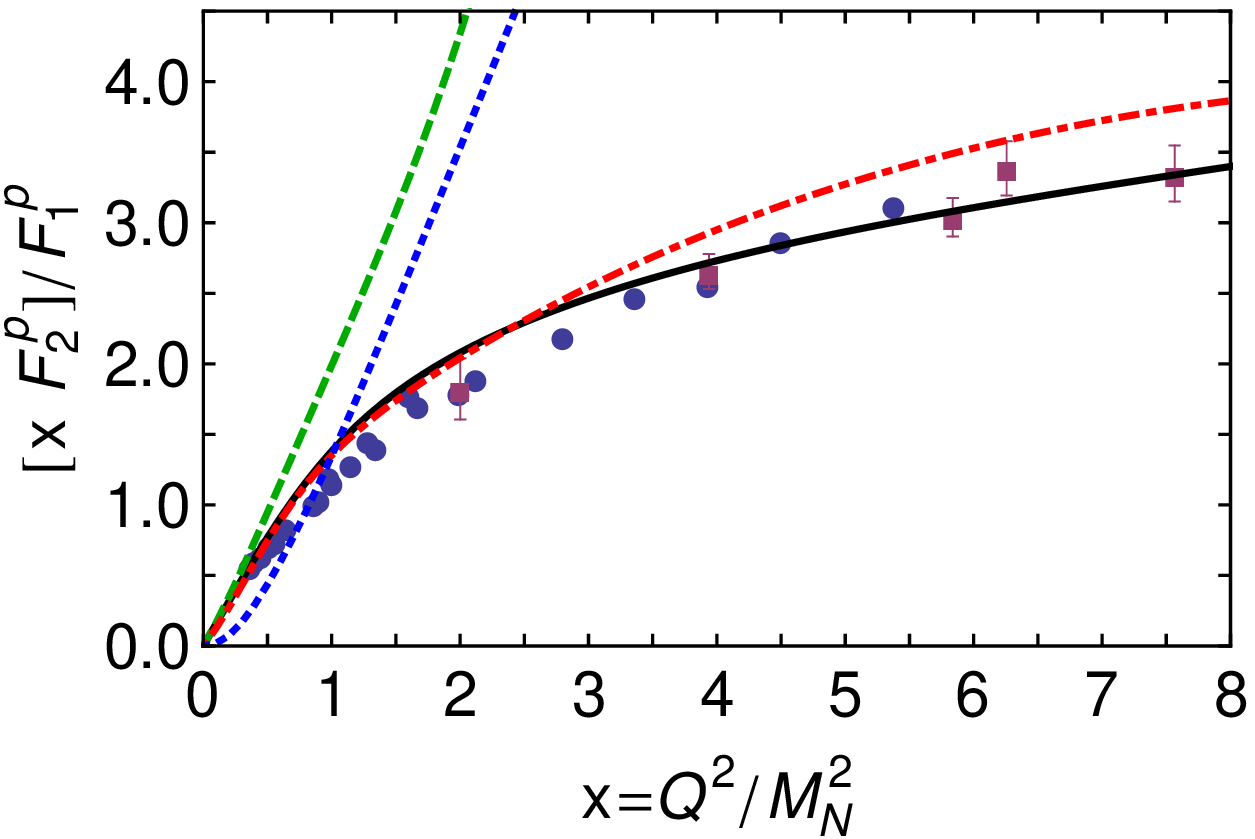}
\hspace*{0.40cm}
\includegraphics[clip,width=0.46\textwidth,height=0.275\textheight]
{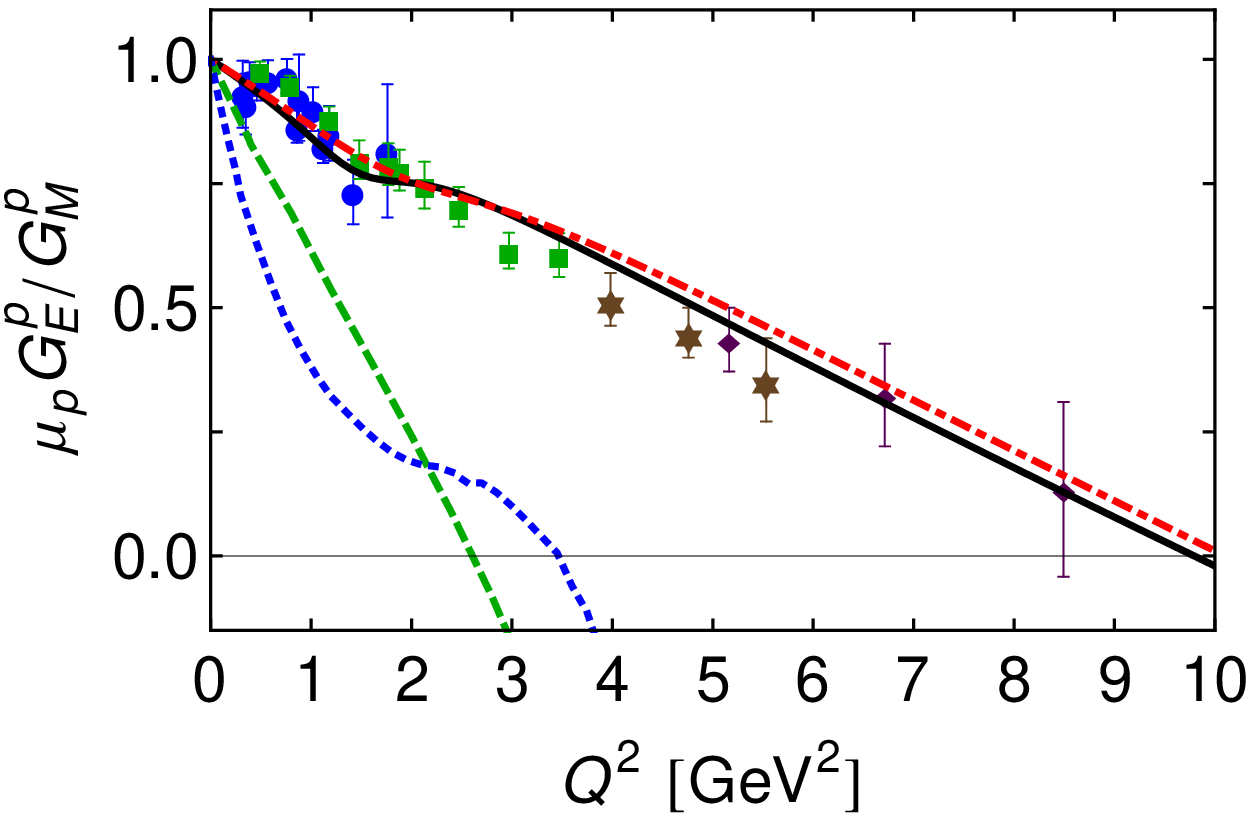}}
\caption{\label{fig:NucNuc_Ratios} 
{\it Left panel:} Proton ratio $R_{21}(x) = x F_2(x)/F_1(x)$, $x=Q^2/M_N^2$. 
Curves: solid (black) -- full result, determined from the complete proton 
Faddeev wave function and current; dot-dashed (red) -- momentum-dependence of 
scalar-diquark contribution; dashed (green) -- momentum-dependence produced by 
that piece of the scalar diquark contribution to the proton's Faddeev wave 
function which is purely $S$-wave in the rest-frame; dotted (blue) -- 
momentum-dependence of pseudovector diquark contribution. All partial 
contributions have been renormalised to produce unity at $x=0$. Experimental 
data taken from Ref.~\protect\cite{Cates:2011pz}.
{\it Right panel:} Computed ratio of proton electric and magnetic Sachs form 
factors. The legend for the curves is the same than that of the lower-left 
panel.
Data:
circles (blue)~\protect\cite{Gayou:2001qt};
squares (green)~\protect\cite{Punjabi:2005wq};
asterisks (brown)~\protect\cite{Puckett:2010ac};
and diamonds (purple)~\protect\cite{Puckett:2011xg}.
}
\end{figure}

Consider now the ratio $R_{21}(x) = x F_2(x)/F_1(x)$ of the proton. A clear 
conclusion from the left panel of Fig.~\ref{fig:NucNuc_Ratios} is that 
pseudovector diquark correlations have little influence on the momentum 
dependence of $R_{21}(x)$. Their contribution is indicated by the dotted 
(blue) curve, which was obtained by setting the scalar diquark component of the 
proton's Faddeev amplitude to zero and renormalising the result to unity at 
$x=0$. As apparent from the dot-dashed (red) curve, the evolution of 
$R_{21}(x)$ with $x$ is primarily determined by the proton's scalar 
diquark component. As we have explained above, in this component, the valence 
$d$-quark is sequestered inside the soft scalar diquark correlation so that the 
only objects within the nucleon which can participate in a hard scattering event 
are the valence $u$-quarks. The scattering from the proton's valence $u$-quarks 
is responsible for the momentum dependence of $R_{21}(x)$. However, the dashed 
(green) curve in the left panel of Fig.~\ref{fig:NucNuc_Ratios} reveals 
something more, {\it i.e.} components of the nucleon associated with 
quark-diquark orbital angular momentum $L\geq1$ in the nucleon rest frame are 
critical in explaining the data. Notably, the presence of such components is an 
inescapable consequence of the self-consistent solution of a realistic 
Poincar\'e-covariant Faddeev equation for the nucleon. 

It is natural now to consider the ratio of the proton electric and 
magnetic Sachs form factors: $R_{EM}(Q^2) = \mu_p G_E(Q^2)/G_M(Q^2)$, 
$\mu_p=G_M(0)$, drawn in the right panel of Fig.~\ref{fig:NucNuc_Ratios}. As 
with $R_{21}$, the momentum dependence of $R_{EM}(Q^2)$ is principally 
determined by the scalar diquark component of the proton. Moreover, the 
rest-frame $L\geq1$ terms are again seen to be critical in explaining the data: 
the behaviour of the dashed (green) curve highlights the impact of omitting 
these components.


\begin{figure}[!t]
\begin{minipage}[t]{\textwidth}
\begin{minipage}{0.49\textwidth}
\centerline{\includegraphics[clip,width=0.85\textwidth,height=0.275\textheight]
{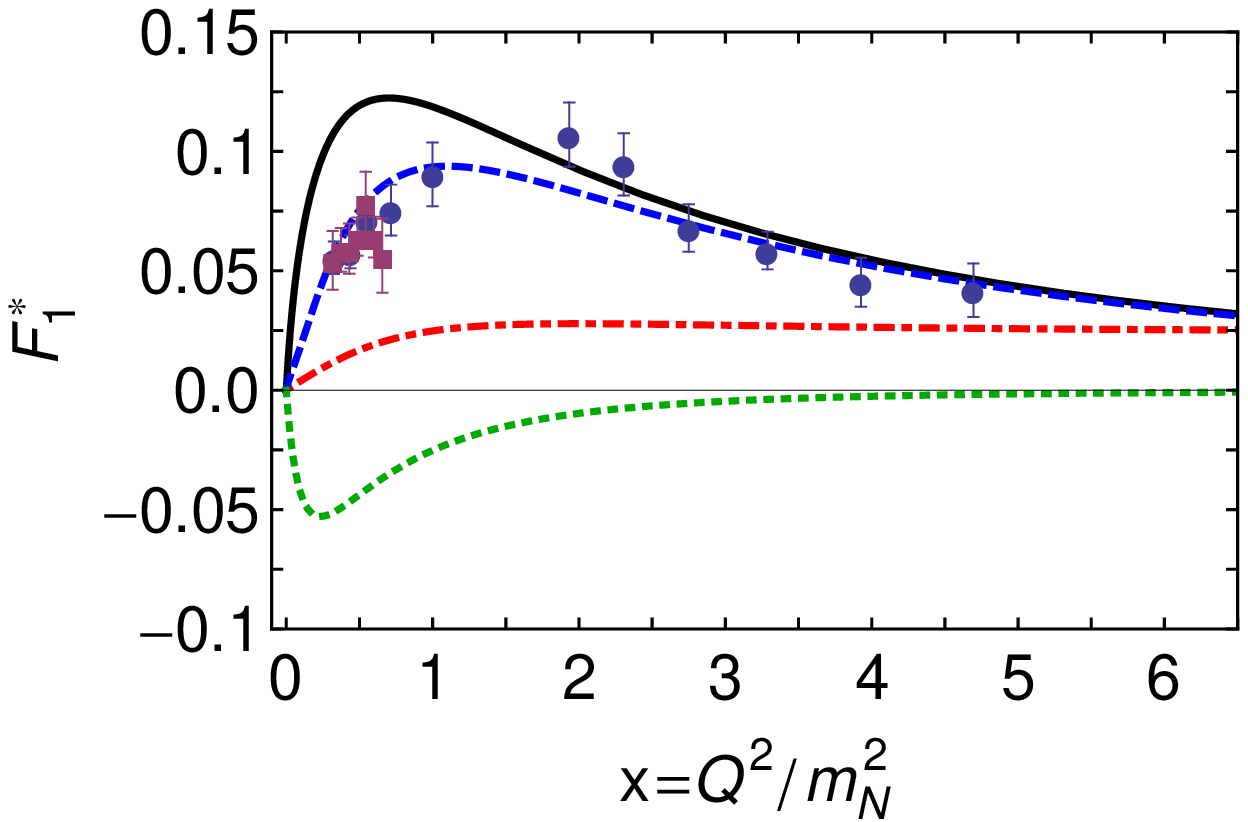}}
\end{minipage}
\begin{minipage}{0.49\textwidth}
\centerline{\includegraphics[clip,width=0.85\textwidth,height=0.275\textheight]
{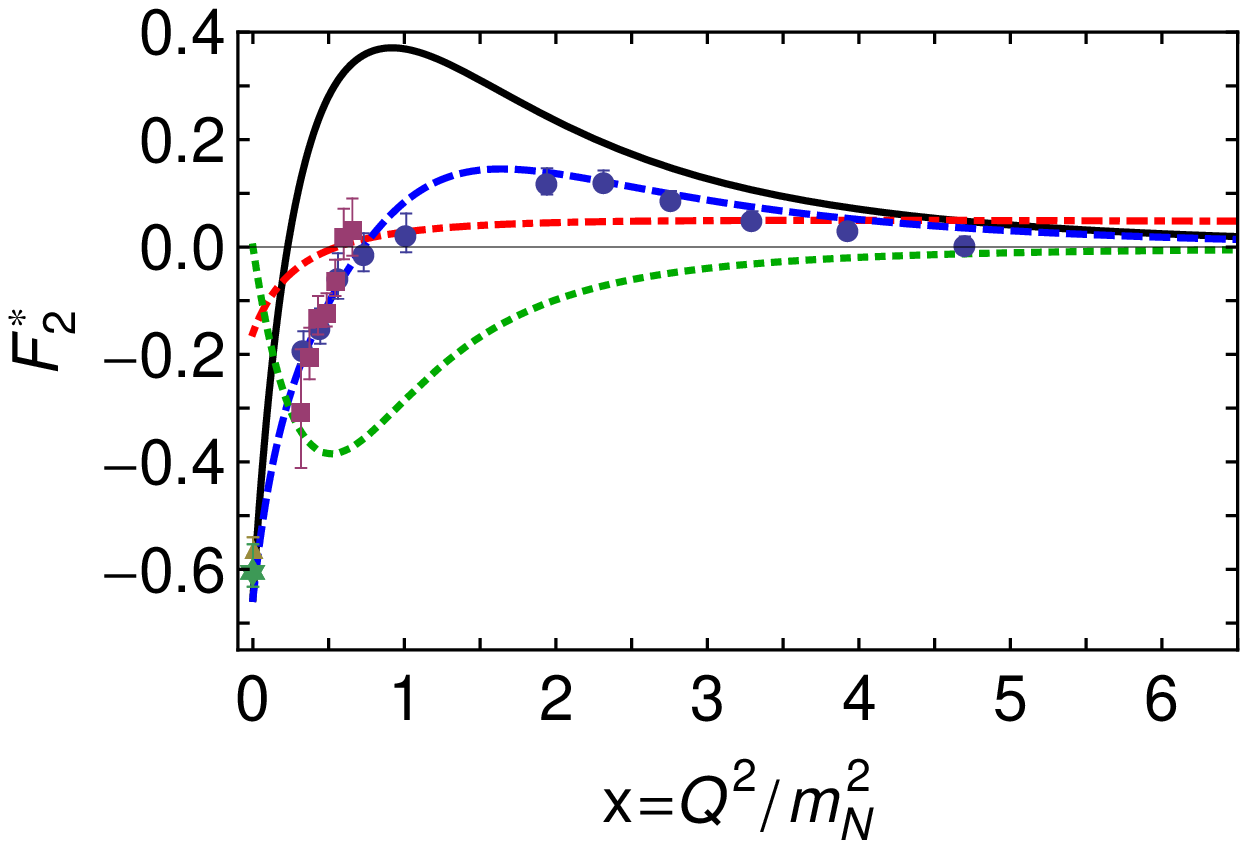}}
\end{minipage}
\end{minipage}
\caption{\label{fig:NucRop_total} \emph{Left panel} -- Dirac transition form 
factor $F_{1}^{\ast}(x)$ as a function of $x=Q^2/m_N^2$ with $Q$ the photon's 
momentum and $m_{N}$ the nucleon's mass. Solid (black) curve, QCD-kindred 
prediction; dot-dashed (red) curve, contact-interaction result; dotted (green) 
curve, inferred meson-cloud contribution; and dashed (blue) curve, anticipated 
complete result. \emph{Right panel} -- Pauli transition form factor, 
$F_{2}^{\ast}(x)$, with same legend. Data in both panels: circles 
(blue)~\cite{Aznauryan:2009mx}; triangle (gold)~\cite{Dugger:2009pn}; squares 
(purple)~\cite{Mokeev:2012vsa}; and star (green)~\cite{Agashe:2014kda}.}
\vspace*{-0.50cm}
\end{figure}

\section{The \mbox{\boldmath $\gamma^\ast N \to N(1440)$} Transition}
\label{subsec:FFnucrop}

Jefferson Lab experiments~\cite{Aznauryan:2011qj, Aznauryan:2009mx, 
Dugger:2009pn, Mokeev:2012vsa} have yielded precise nucleon-to-Roper transition 
form factors and thereby exposed the first zero seen in any hadron form 
factor or transition amplitude. We have performed in 
Ref.~\cite{Segovia:2015hra} the first continuum treatment of this problem using 
the power of relativistic quantum field theory. It is such calculation which 
has allowed us to state that the Roper resonance appears to be the nucleon's 
first radial excitation whose unexpectedly low mass arise from a dressed-quark 
core that is shielded by a meson-cloud.

The left panel of Fig.~\ref{fig:NucRop_total} depicts the Dirac transition form 
factor, $F_{1}^{\ast}$, which vanishes at $x=0$ owing to orthogonality between 
the proton and its radial excitation. Our calculation, solid (black) curve, 
agrees quantitatively in magnitude and qualitatively in trend with the data on 
$x\gtrsim 2$. The nature of our prediction owes fundamentally to the 
QCD-derived momentum-dependence of the propagators and vertices employed in 
formulating the bound-state and scattering problems. This point is further 
highlighted by the contact-interaction result (dot-dashed-red): with 
momentum-independent propagators and vertices, the prediction disagrees both 
quantitatively and qualitatively with the data. Experiment is evidently a 
sensitive tool with which to chart the nature of the quark-quark interaction and 
hence discriminate between competing theoretical hypotheses; and it is plainly 
settling upon an interaction that produces the momentum-dependent dressed-quark 
mass which characterises QCD \cite{Bowman:2005vx, Bhagwat:2006tu, 
Roberts:2007ji}.

The mismatch between our prediction and the data on $x\lesssim 2$ is also 
revealing. As seen previously, \emph{e.g}.\ Refs.~\cite{Cloet:2008re, 
Segovia:2014aza, Roberts:2015dea}, this is the domain upon which meson-cloud 
contributions are expected to be important. An estimate of that contribution is 
provided by the dotted (green) curve in the left panel of 
Fig.~\ref{fig:NucRop_total}. If this curve is added to our prediction, then one 
obtains the dashed (blue) curve, which is a least-squares fit to the data on 
$x\in [0,5]$. The correction curve has fallen to just $20\%$ of its maximum 
value by $x=2$ and vanishes rapidly thereafter so that our prediction alone 
remains as the explanation of the data.

The right panel of Fig.~\ref{fig:NucRop_total} depicts the Pauli transition 
form factor, $F_{2}^{\ast}$. All observations made regarding $F_{1}^{\ast}$ 
also apply here, including those concerning the estimated meson-cloud 
contributions. Importantly, the existence of a zero in $F_{2}^{\ast}$, a 
prominent feature of the data, is not influenced by meson-cloud effects, 
although its precise location is. Thus any realistic approach to the 
proton-Roper transition must describe a zero in $F_{2}^{\ast}$. It is worth 
noting in addition that our prediction $F_{2}^{\ast}(x=0)=-0.65$, \emph{i.e}.\ 
for the Pauli form factor at the photoproduction point, is consistent with 
contemporary experiment: $-0.58 \pm 0.02$~\cite{Dugger:2009pn} and $-0.62 \pm 
0.04$~\cite{Agashe:2014kda}.

In two separate ways, the electromagnetically induced $N\to R$ transition can 
be considered as a sum of three distinct terms~\cite{Segovia:2014aza}:
\begin{itemize}
\item {\bf T1 = diquark dissection.} \emph{T1A} -- sca\-lar diquark 
in both the initial- and final-state baryon, \emph{T1B} -- pseudovector diquark 
in both the initial- and final-state baryon, and \emph{T1C} -- a different 
diquark in the initial- and final-state baryon.
\item {\bf T2 = scatterer dissection.} \emph{T2A} -- photon strikes 
a bystander dressed-quark, \emph{T2B} -- photon interacts with a diquark, 
elastically or causing a transition scalar\,$\leftrightarrow$\,pseudovector, 
and \emph{T2C} -- photon strikes a dressed-quark in-flight, as one diquark 
breaks up and another is formed, or appears in one of the two associated 
``seagull'' terms.
\end{itemize}

\begin{figure}[t]
\begin{minipage}{0.49\textwidth}
\centerline{%
\includegraphics[clip,width=0.85\textwidth,height=0.25\textheight]
{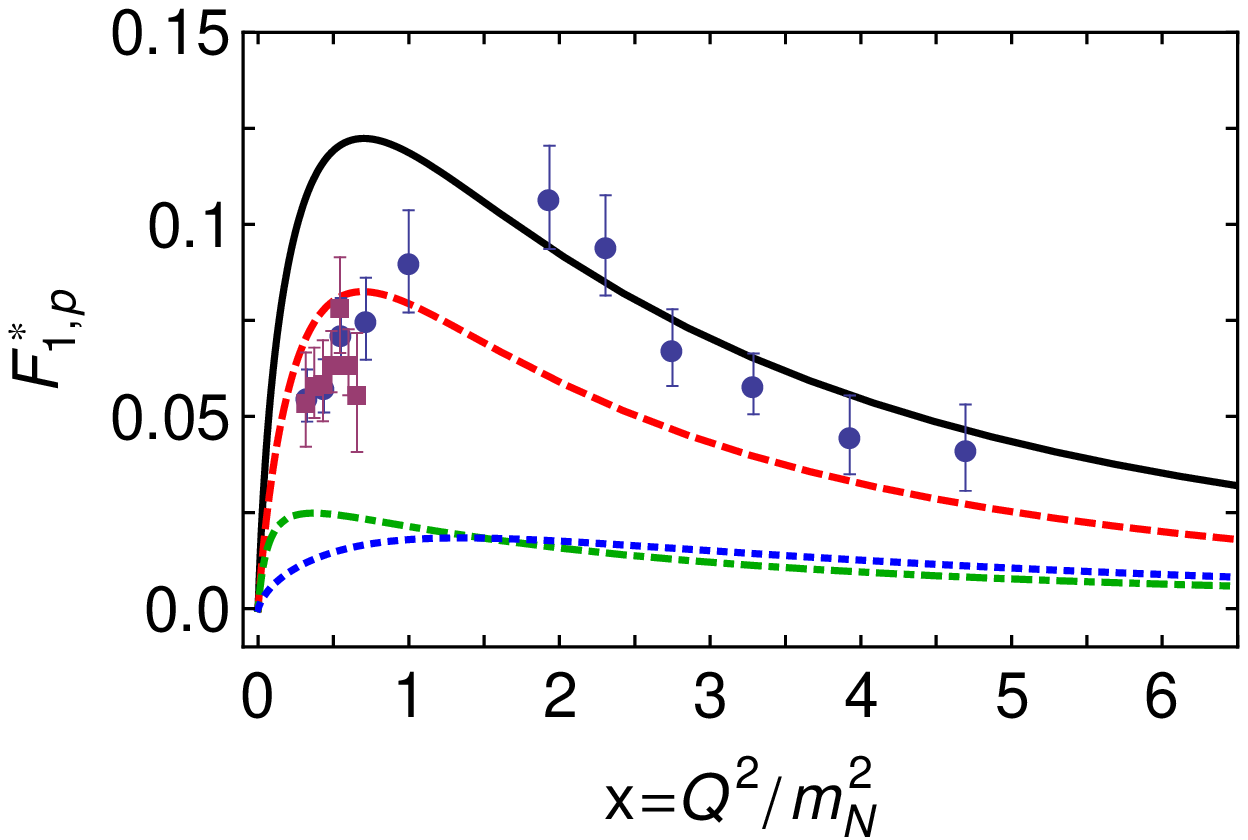}}
\vspace*{0.0ex}
\centerline{%
\includegraphics[clip,width=0.85\textwidth,height=0.25\textheight]
{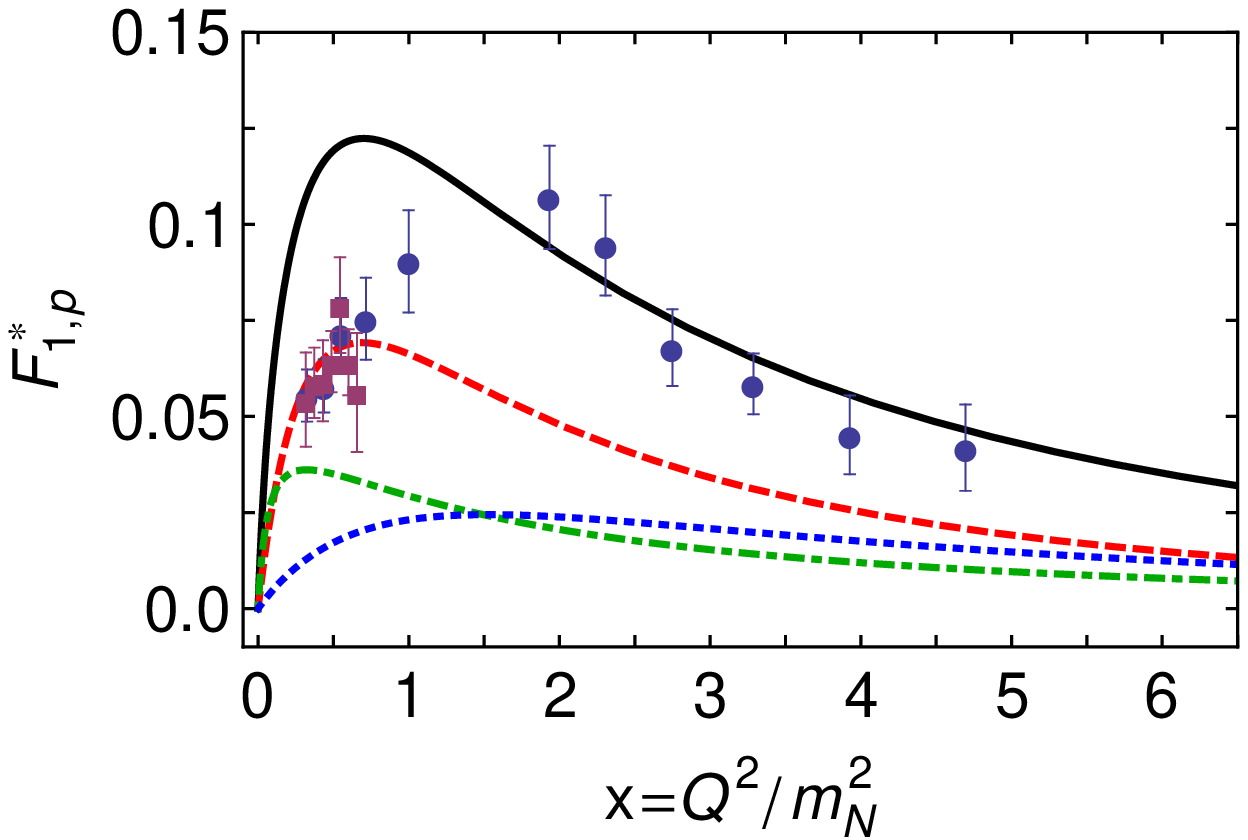}}
\vspace*{0.0ex}
\end{minipage}
\begin{minipage}{0.49\textwidth}
\centerline{%
\includegraphics[clip,width=0.85\textwidth,height=0.25\textheight]
{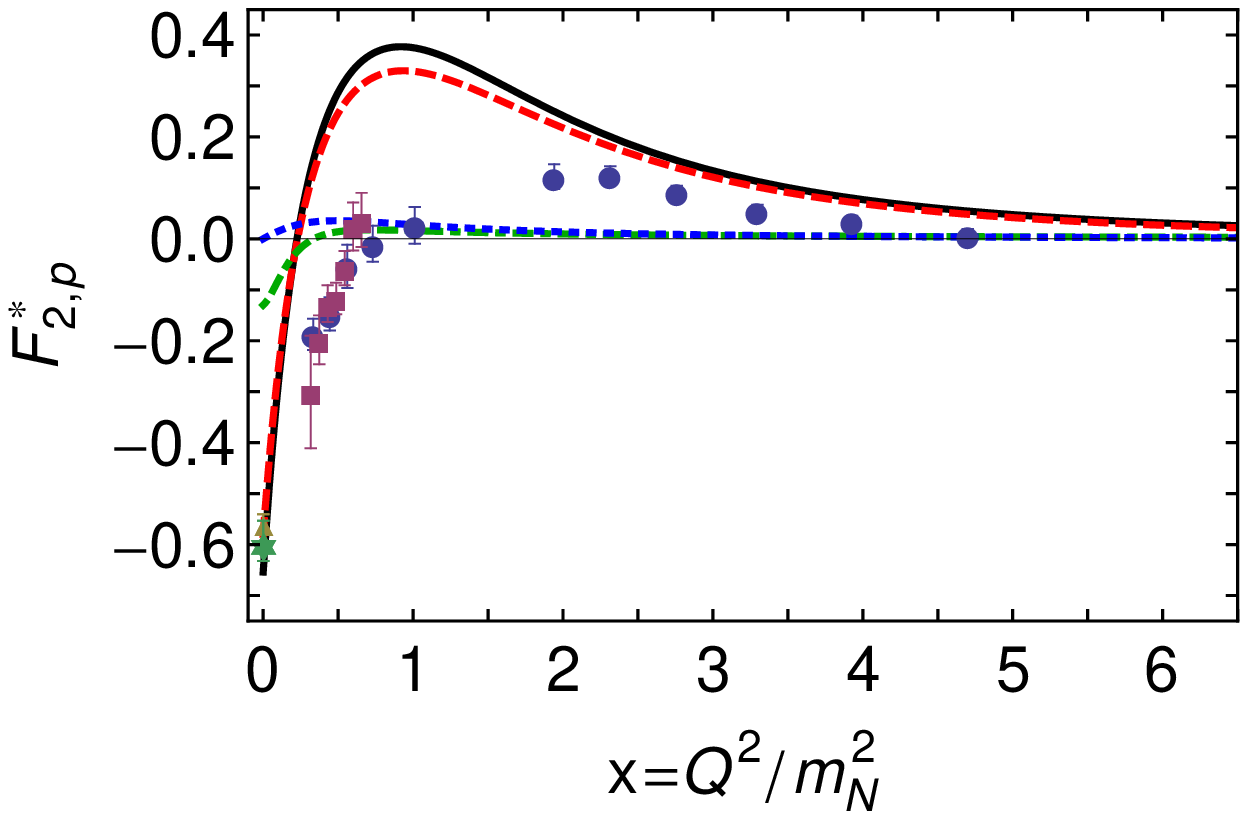}}
\vspace*{0.0ex}
\centerline{%
\includegraphics[clip,width=0.85\textwidth,height=0.25\textheight]
{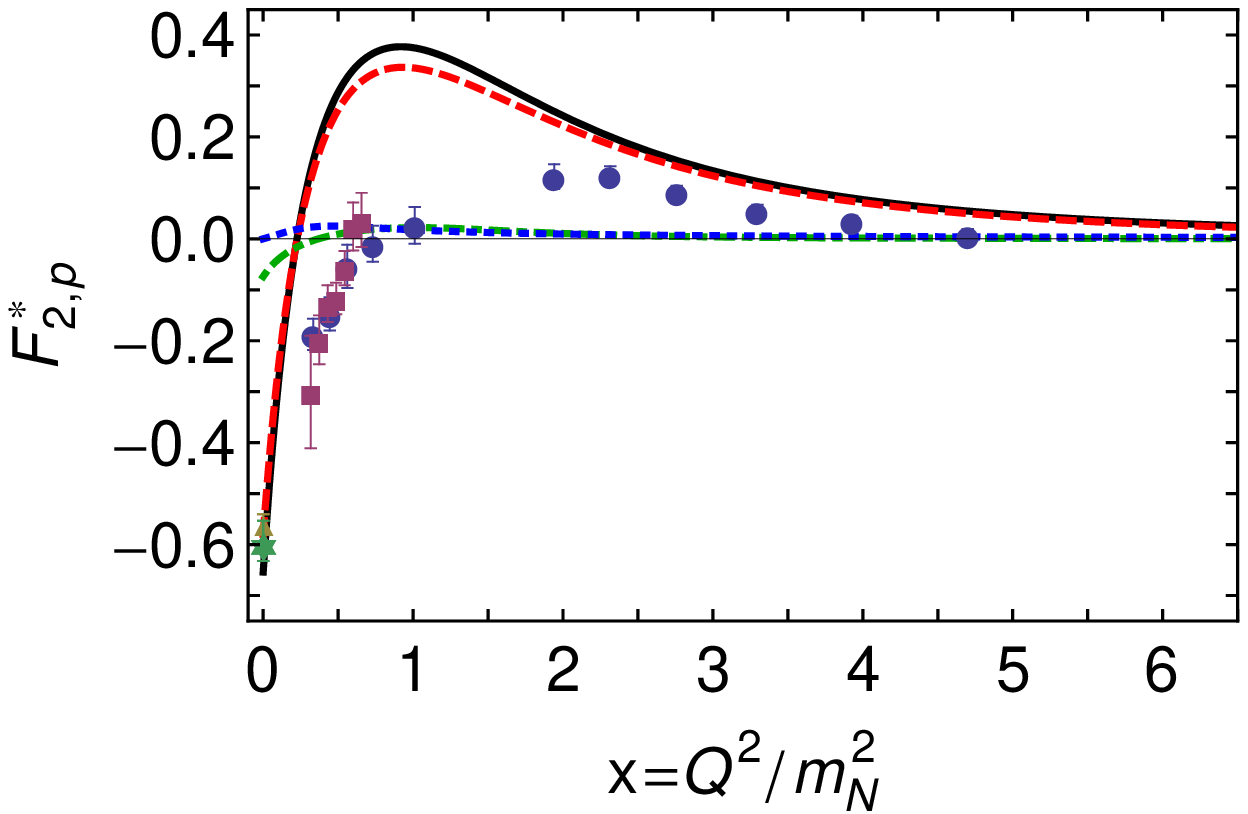}}
\end{minipage}
\caption{\label{fig:NucRop_Disecction} \emph{Left panels (Right panels)} -- The 
$\gamma^{\ast}\,p\to R^+$ Dirac (Pauli) transition form factor, 
$F_{1,p}^{\ast}$ ($F_{2,p}^{\ast}$), as a function of $x=Q^2/m_N^2$, computed 
as described in Ref.~\cite{Segovia:2015hra} (solid black curve).
\emph{Upper panels} -- diquark breakdown: \emph{T1A} (dashed red), scalar 
diquark in both $p$, $R^+$; \emph{T1B} (dot-dashed green), pseudovector diquark 
in both $p$, $R^+$; \emph{T1C} (dotted blue), scalar diquark in $p$, 
pseudovector diquark in $R^+$, and vice versa.
\emph{Lower panels} -- scatterer breakdown: \emph{T2A} (red dashed), 
photon strikes an uncorrelated dressed quark; \emph{T2B} (dot-dashed green), 
photon strikes a diquark; and \emph{T2C} (dotted blue), diquark breakup 
contributions, including photon striking exchanged dressed-quark.
Data:
circles (blue)~\cite{Aznauryan:2009mx};
squares (purple)~\cite{Mokeev:2012vsa};
triangle (gold)~\cite{Dugger:2009pn};
and star (green)~\cite{Agashe:2014kda}.
}
\end{figure}

The anatomy of the $\gamma\,p\to R^+$ Dirac and Pauli transition form factors 
is revealed when combining the information provided by the T1 and T2 
dissections, see Fig.~\ref{fig:NucRop_Disecction}. The Dirac form factor is 
primarily given by the contribution in which a photon strikes a bystander 
dressed quark that is partnered by a scalar-diquark: $[ud]$. However, all other 
processes have non-negligible contributions. In exhibiting these features, 
$F_{1,p}^{\ast}$ shows marked qualitative similarities to the proton's elastic 
Dirac form factor (\emph{cf}. Fig.~3 in Ref.~\cite{Cloet:2008re}).

In the case of the Pauli transition form factor, a single contribution is 
overwhelmingly important, \emph{viz}.\ photon strikes a bystander dressed-quark 
in association with $[ud]$ in the proton and $R^+$. No other diagram makes a 
significant contribution. A comparison with Fig.~4 in Ref.~\cite{Cloet:2008re} 
reveals that the same may be said for the dressed-quark core component of the 
proton's elastic Pauli form factor.

Given that the diquark content of the proton and $R^+$ are almost identical, 
with the $\psi_0 \sim u[ud]$ component contributing roughly 60\% of the charge 
of both systems, the qualitative similarity between the proton elastic and 
proton-Roper transition form factors is not surprising. This observation 
immediately raises the issue of whether and how that similarity is transmitted 
into the flavour separated form factors.

If one supposes that $s$-quark contributions to the nucleon-Roper transitions 
are negligible, as is the case for nucleon elastic form factors, and assumes 
isospin symmetry, then a flavour separation of the transition form factors is 
accomplished by combining results for the $\gamma^{\ast}\,p\to\, R^+$ and 
$\gamma^{\ast}\,n\to 
R^0$ transitions:
\begin{equation}
\label{FlavourSep}
F_{1(2),u}^{\ast} = 2 F_{1(2)}^{\ast,p} + F_{1(2)}^{\ast,n}, \quad\quad
F_{1(2),d}^{\ast} = 2 F_{1(2)}^{\ast,n} + F_{1(2)}^{\ast,p},
\end{equation}
where $p$ and $n$ are superscripts that indicate, respectively, the charged and 
neutral nucleon-Roper reactions. Our conventions are that $F_{1(2),u}^{\ast}$ 
and $F_{1(2),d}^{\ast}$ refer to the $u$- and $d$-quark contributions to the 
equivalent Dirac (Pauli) form factors of the $\gamma^{\ast} p\to R^+$ reaction, 
and the results are normalised such that the \emph{elastic} Dirac form factors 
of the proton and charged-Roper yield $F_{1u}(Q^2=0)=2$, $F_{1d}(Q^2=0)=1$, 
thereby ensuring that these functions count $u$- and $d$-quark content in the 
bound-states.


\begin{figure}[t]
\begin{minipage}{0.49\textwidth}
\centerline{%
\includegraphics[clip,width=0.85\textwidth,height=0.25\textheight]
{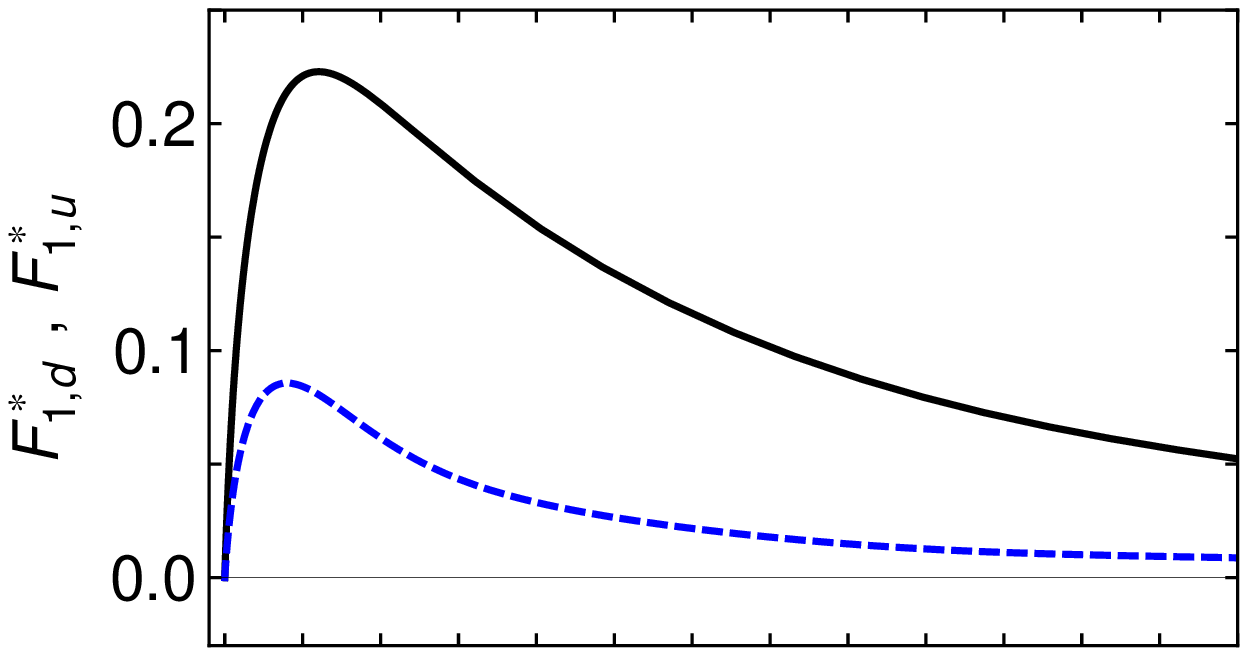}}
\vspace*{-8.0ex}
\centerline{%
\hspace*{-0.36cm}
\includegraphics[clip,width=0.89\textwidth,height=0.25\textheight]
{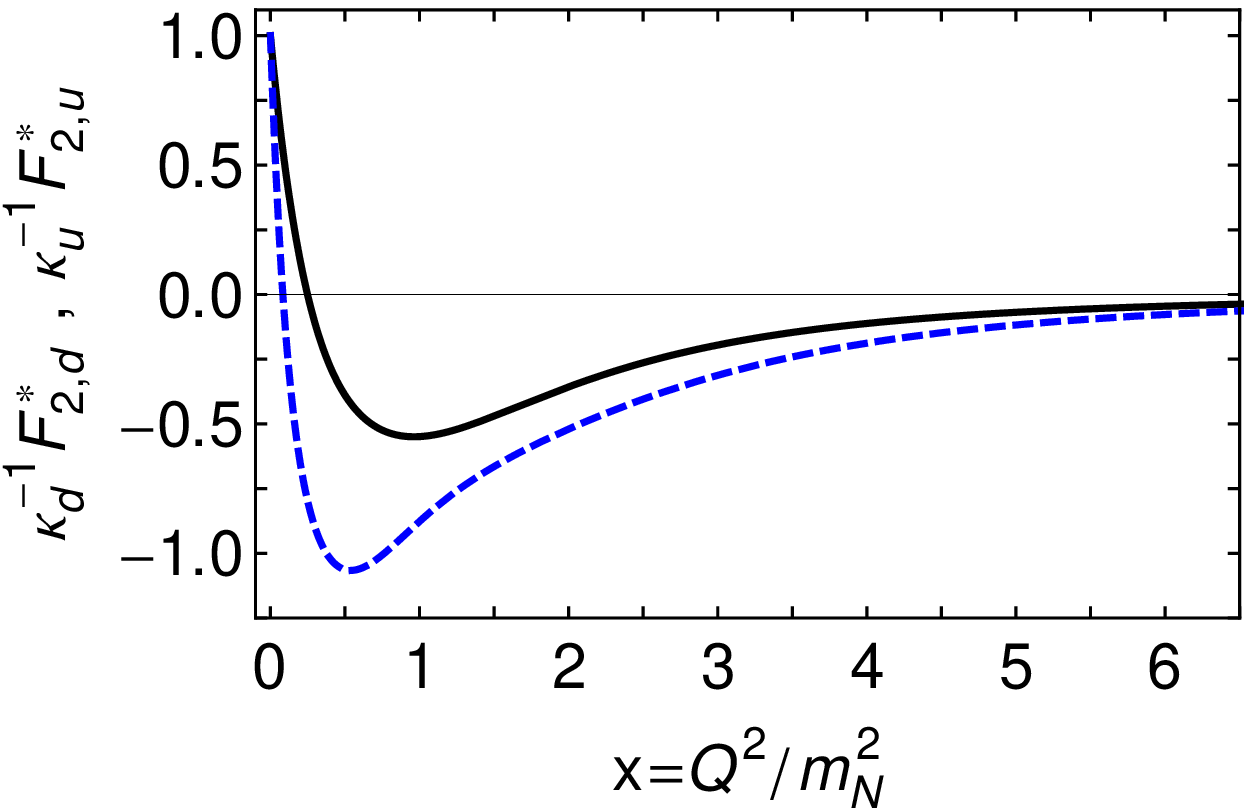}}
\vspace*{0.0ex}
\end{minipage}
\begin{minipage}{0.49\textwidth}
\centerline{%
\includegraphics[clip,width=0.85\textwidth,height=0.25\textheight]
{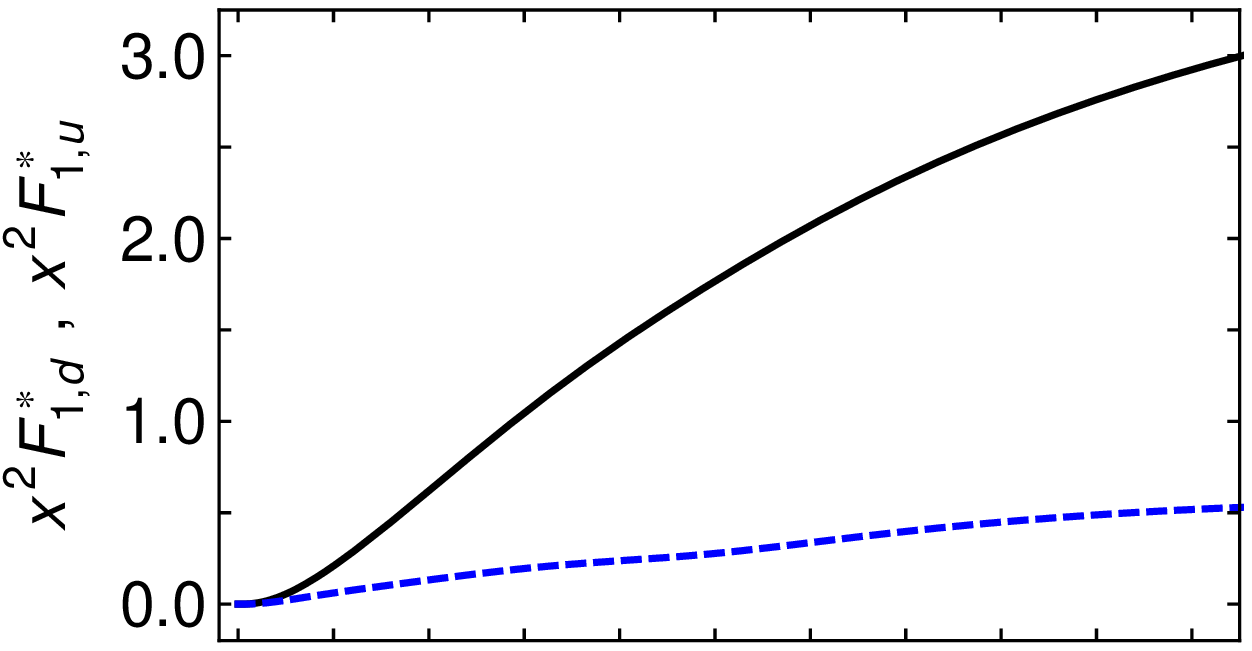}}
\vspace*{-8.0ex}
\centerline{%
\hspace*{-0.32cm}
\includegraphics[clip,width=0.89\textwidth,height=0.25\textheight]
{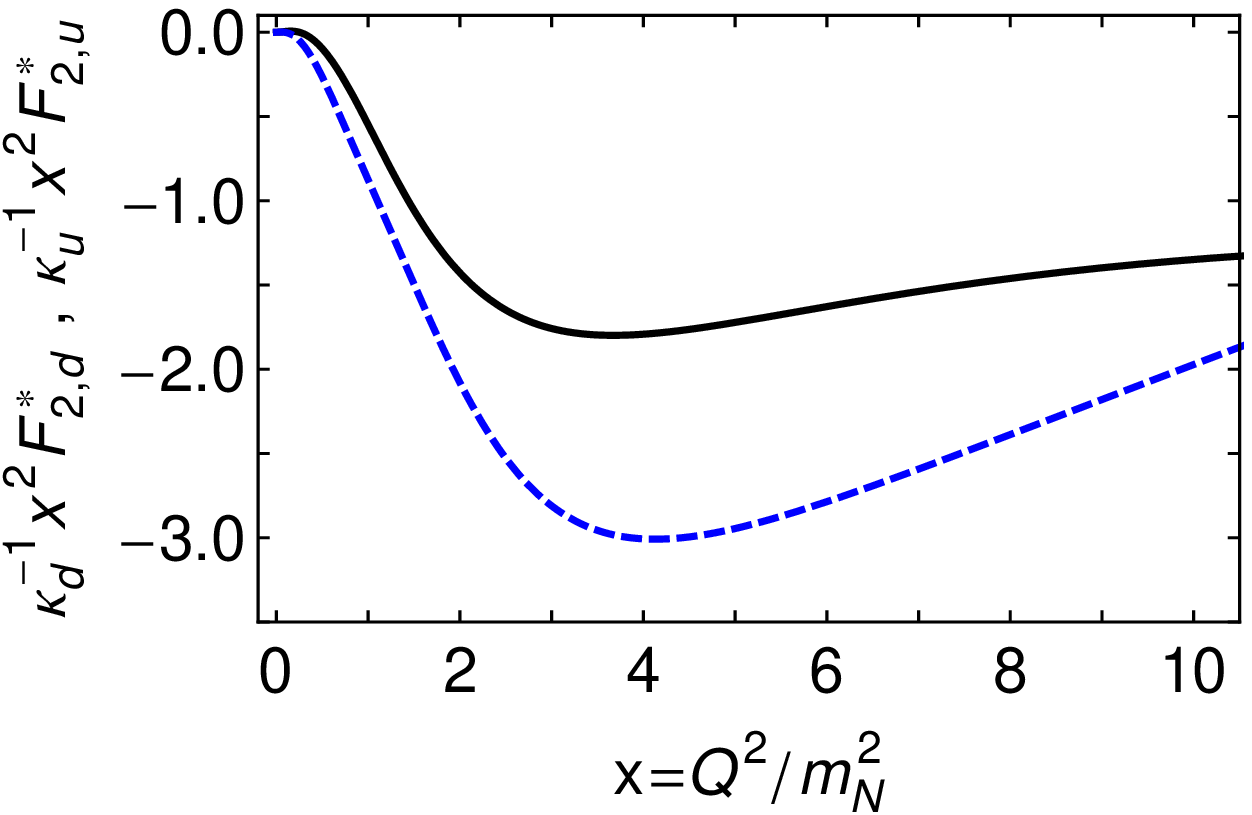}}
\end{minipage}
\caption{\label{fig:NucRop_Flavour} \emph{Left panels} -- Flavour separation 
$\gamma^{\ast}\,p\to R^+$ transition form factors: $u$-quark, solid black; and 
$d$-quark, dashed blue.  \emph{Upper-left panel} -- Dirac transition form 
factor. \emph{Lower-left panel} -- Pauli transition form factor, with 
$\kappa_u^\ast  = F_{2,u}^\ast(0) = -0.91$, $\kappa_d^\ast = F_{2,d}^\ast(0) = 
0.14$.
\emph{Right panels} -- $x^2=(Q^2/m_N^2)^2$-weighted behaviour of the flavour 
separated $\gamma^{\ast}\,p\to R^+$ transition form factors: $u$-quark, solid 
black; and $d$-quark, dashed blue.  \emph{Upper-right (lower-right) panel} -- 
Dirac (Pauli) transition form factor.
}
\end{figure}

The upper panels of Fig.~\ref{fig:NucRop_Flavour}, depicting the 
flavour-separated Dirac transition form factor, show an obvious similarity to 
the analogous form factor determined in elastic scattering: the $d$-quark 
contribution is less-than half the $u$-quark contribution for momenta 
sufficiently far outside the neighbourhood of $x=Q^2/m_N^2=0$ within which they 
both vanish; and the $d$-quark contribution falls more rapidly after their 
almost coincident maxima. The noticeable difference, however, is the absence of 
a zero in $F_{1,d}^{\ast}$, which is a salient feature of the analogous proton 
elastic form factor.

The lower panels of Fig.~\ref{fig:NucRop_Flavour} depict the 
flavour-separated Pauli transition form factor. In this instance the 
similarities are less obvious, but they are revealed once one recognises that 
the rescaling factors satisfy $|\kappa_d^\ast/\kappa_u^\ast | < \tfrac{1}{6}$ 
\emph{cf}.\ a value of $\sim \tfrac{2}{5}$ in the elastic 
case~\cite{Cloet:2008re}. Accounting for this, the behaviour of the $u$- and 
$d$-quark contributions to the charged-Roper Pauli transition form factor are 
comparable with the kindred contributions to the elastic form factor, especially 
insofar as the $d$-quark contribution falls dramatically on $x\gtrsim 4$ whereas 
the $u$-quark contribution evolves more slowly.

An explanation for the pattern of behaviour in Fig.~\ref{fig:NucRop_Flavour} 
is much the same as that for the analogous proton elastic form 
factors because the diquark content of the proton and its first radial 
excitation are almost identical. In both systems, the dominant piece of the 
associated Faddeev wave functions is $\psi_0$, namely a $u$-quark in tandem 
with a $[ud]$ (scalar diquark) correlation, which produces 62\% of each 
bound-state's normalisation~\cite{Segovia:2015hra}. If $\psi_0$ were the sole 
component in both the proton and charged-Roper, then $\gamma$--$d$-quark 
interactions would receive a $1/x$ suppression on $x>1$, because the $d$-quark 
is sequestered in a soft correlation, whereas a spectator $u$-quark is always 
available to participate in a hard interaction. At large-$x$, therefore, scalar 
diquark dominance leads one to expect $F^\ast_d \sim F^\ast_u/x$.  Naturally, 
precise details of this $x$-dependence are influenced by the presence of 
pseudovector diquark correlations in the initial and final states, which 
guarantees that the singly-represented quark, too, can participate in a hard 
scattering event, but to a lesser extent.

The infrared behaviour of the flavour-separated $\gamma^{\ast} p \to R^+$ 
transition form factors owes to a complicated interference between the 
influences of orthogonality, which forces $F^\ast_{1,u}(0)=0=F^\ast_{1,d}(0)$, 
and quark-core and MB\,FSI contributions. However, whilst the latter pair act 
in similar ways for both elastic and transition form factors, orthogonality is 
a fundamental difference between the two processes and it is therefore likely 
to be the dominant effect at infrared momenta.

The information contained in Figs.~\ref{fig:NucRop_Disecction} 
and~\ref{fig:NucRop_Flavour} provides clear evidence in support of the notion 
that many features in the measured behaviour of $\gamma^{\ast} N \to R$ 
electromagnetic transition form factors are primarily driven by the presence of 
strong diquark correlations in the nucleon and its first radial excitation. In 
our view, inclusion of a ``meson cloud'' cannot qualitatively affect the 
salient features of these transition form factors, any more than it does the 
analogous nucleon elastic form factors~\cite{Cloet:2012cy, Cloet:2014rja}.


\section{Conclusions}
\label{sec:conclusions}

We have explained how the emergent phenomenon of dynamical chiral symmetry 
breaking ensures that Poincar\'e covariant analyses of the three valence-quark 
scattering problem in continuum quantum field theory yield a picture in which 
binding arises primarily through the sum of two separate contributions. One 
involves aspects of the non-Abelian character of QCD that are expressed in the 
strong running coupling and generate tight, dynamical colour-antitriplet 
quark-quark correlations. This attraction is magnified by quark exchange 
associated with diquark breakup and reformation, which is required in order to 
ensure that each valence-quark participates in all diquark correlations to the 
complete extent allowed by its quantum numbers.

The existence of strong diquark correlations inside the nucleon and its 
resonances has numerous observable consequences and we have presented herein 
some of them related with the nucleon's elastic and nucleon-to-Roper transition 
electromagnetic form factors. 

In the case of the nucleon, we have seen that Poincar\'e covariance, which 
demands the presence of dressed-quark orbital angular momentum in the nucleon, 
and dominant scalar diquark correlations are sufficient to understand empirical 
extractions of the flavour-separated Dirac and Pauli form factors, and the 
$Q^2$-behaviour of the proton's electromagnetic ratios.

The anatomy of the $\gamma^{\ast}\,p\to R^+$ Dirac and Pauli transition form 
factors has been presented. We concluded that the contribution in which the 
photon strikes a bystander $u$-quark in association with a scalar diquark 
$[ud]$ is dominant for both Dirac and Pauli transition form factors. Moreover, 
their flavour separated versions show similarities to the analogous form 
factors determined in elastic scattering. In the case of Dirac transition form 
factor, the $d$-quark contribution is less-than half the $u$-quark contribution 
for momenta sufficiently far outside the neighbourhood of $Q^2=0$; and the 
$d$-quark contribution falls more rapidly at very large momenta. The 
similarities are less obvious in the case of the Pauli transition form factor 
but one can appreciate that the $d$-quark contribution falls dramatically at 
large momenta whereas the $u$-quark contribution evolves more slowly.


\begin{acknowledgement}
The material described in this contribution is drawn from work completed in 
collaboration with numerous ex\-ce\-llent people, to all of whom we are greatly 
indebted.
We thank R.~Gothe, V.~Mokeev and V.~Burkert for suggesting this problem, and 
T.S.-H.\,Lee and T.\,Sato for numerous informative discussions.
This work was supported by: the Alexander von Humboldt Foundation; and the U.S. 
Department of Energy, Office of Science, Office of Nuclear Physics, under 
contract no. DE-AC02-06CH11357.
\end{acknowledgement}


\bibliography{SEGOVIA_Jorge_CONF12}

\begin{thebibliography}{30}

\bibitem{Cloet:2013jya}
I.C. Cloët, C.D. Roberts, Prog. Part. Nucl. Phys. \textbf{77}, 1 (2014)

\bibitem{Segovia:2015ufa}
J.~Segovia, C.D. Roberts, S.M. Schmidt, Phys. Lett. \textbf{B750}, 100 (2015)

\bibitem{Alexandrou:2006cq}
C.~Alexandrou, P.~de~Forcrand, B.~Lucini, Phys. Rev. Lett. \textbf{97}, 222002
  (2006)

\bibitem{Babich:2007ah}
R.~Babich, N.~Garron, C.~Hoelbling, J.~Howard, L.~Lellouch, C.~Rebbi, Phys.
  Rev. \textbf{D76}, 074021 (2007)

\bibitem{Eichmann:2016yit}
G.~Eichmann, H.~Sanchis-Alepuz, R.~Williams, R.~Alkofer, C.S. Fischer, Prog.
  Part. Nucl. Phys. \textbf{91}, 1 (2016)

\bibitem{Eichmann:2016hgl}
G.~Eichmann, C.S. Fischer, H.~Sanchis-Alepuz (2016),
  \texttt{arXiv:hep-ph/1607.05748}

\bibitem{Segovia:2016zyc}
J.~Segovia, C.D. Roberts, Phys. Rev. \textbf{C94}, 042201 (2016)

\bibitem{Segovia:2013rca}
J.~Segovia, C.~Chen, C.D. Roberts, S.~Wan, Phys. Rev. \textbf{C88}, 032201
  (2013)

\bibitem{Segovia:2013uga}
J.~Segovia, C.~Chen, I.C. Cloët, C.D. Roberts, S.M. Schmidt, S.~Wan, Few Body
  Syst. \textbf{55}, 1 (2014)

\bibitem{Segovia:2014aza}
J.~Segovia, I.C. Cloët, C.D. Roberts, S.M. Schmidt, Few Body Syst.
  \textbf{55}, 1185 (2014)

\bibitem{Segovia:2015hra}
J.~Segovia, B.~El-Bennich, E.~Rojas, I.C. Cloët, C.D. Roberts, S.S. Xu, H.S.
  Zong, Phys. Rev. Lett. \textbf{115}, 171801 (2015)

\bibitem{Eichmann:2009qa}
G.~Eichmann, R.~Alkofer, A.~Krassnigg, D.~Nicmorus, Phys. Rev. Lett.
  \textbf{104}, 201601 (2010)

\bibitem{Wilson:2011aa}
D.J. Wilson, I.C. Cloet, L.~Chang, C.D. Roberts, Phys. Rev. \textbf{C85},
  025205 (2012)

\bibitem{Cloet:2014rja}
I.C. Cloët, W.~Bentz, A.W. Thomas, Phys. Rev. \textbf{C90}, 045202 (2014)

\bibitem{Cates:2011pz}
G.D. Cates, C.W. de~Jager, S.~Riordan, B.~Wojtsekhowski, Phys. Rev. Lett.
  \textbf{106}, 252003 (2011)

\bibitem{Gayou:2001qt}
O.~Gayou et~al., Phys. Rev. \textbf{C64}, 038202 (2001)

\bibitem{Punjabi:2005wq}
V.~Punjabi et~al., Phys. Rev. \textbf{C71}, 055202 (2005), [Erratum: Phys. Rev.
  C71, 069902(2005)]

\bibitem{Puckett:2010ac}
A.J.R. Puckett et~al., Phys. Rev. Lett. \textbf{104}, 242301 (2010)

\bibitem{Puckett:2011xg}
A.J.R. Puckett et~al., Phys. Rev. \textbf{C85}, 045203 (2012)

\bibitem{Aznauryan:2009mx}
I.G. Aznauryan et~al. (CLAS), Phys. Rev. \textbf{C80}, 055203 (2009)

\bibitem{Dugger:2009pn}
M.~Dugger et~al. (CLAS), Phys. Rev. \textbf{C79}, 065206 (2009)

\bibitem{Mokeev:2012vsa}
V.I. Mokeev et~al. (CLAS), Phys. Rev. \textbf{C86}, 035203 (2012)

\bibitem{Agashe:2014kda}
K.A. Olive et~al. (Particle Data Group), Chin. Phys. \textbf{C38}, 090001
  (2014)

\bibitem{Aznauryan:2011qj}
I.G. Aznauryan, V.D. Burkert, Prog. Part. Nucl. Phys. \textbf{67}, 1 (2012)

\bibitem{Bowman:2005vx}
P.O. Bowman, U.M. Heller, D.B. Leinweber, M.B. Parappilly, A.G. Williams, J.b.
  Zhang, Phys. Rev. \textbf{D71}, 054507 (2005)

\bibitem{Bhagwat:2006tu}
M.S. Bhagwat, P.C. Tandy, AIP Conf. Proc. \textbf{842}, 225 (2006)

\bibitem{Roberts:2007ji}
C.D. Roberts, Prog. Part. Nucl. Phys. \textbf{61}, 50 (2008)

\bibitem{Cloet:2008re}
I.C. Cloët, G.~Eichmann, B.~El-Bennich, T.~Klahn, C.D. Roberts, Few Body Syst.
  \textbf{46}, 1 (2009)

\bibitem{Roberts:2015dea}
C.D. Roberts, J. Phys. Conf. Ser. \textbf{630}, 012051 (2015)

\bibitem{Cloet:2012cy}
I.C. Cloët, G.A. Miller, Phys. Rev. \textbf{C86}, 015208 (2012)

\end{thebibliography}

%
%
%

\end{document}